\documentclass[12pt]{article}
\usepackage{amsfonts}
\usepackage{amsmath}
\usepackage{amssymb}
{\baselineskip 12pt}  
\begin{document}
\title{
\vspace*{0pt} On the scalar product of short and long living
states of neutral kaons\\ in the CPT invariant system}
\author{ \hfill \\ K. Urbanowski\footnote{e--mail:
K.Urbanowski@if.uz.zgora.pl; K.Urbanowski@proton.if.uz.zgora.pl} \\
\hfill  \\
University of Zielona Gora, Institute of Physics, \\
ul. Prof. Z. Szafrana 4a, 65-516 Zielona Gora, Poland.
\vspace*{-10pt}} \maketitle {\noindent}{\em PACS numbers:}
03.65.Ca., 11.30.Er.,
11.10.St., 14.40.Aq. 

\begin{abstract}
This paper contains a detailed analysis  of the properties of the
scalar product of short and long living superpositions of neutral
$|K_{0}\rangle$, $|{\overline{K}}_{0}\rangle$ mesons. It is shown
for the exact effective Hamiltonian for neutral meson subsystem
that the scalar product of its eigenvectors, which correspond with
these short and long living superpositions, can not be real with
the assumption of  CPT conserved and CP violated. The standard
conclusion obtained within the Lee-Oehme-Yang theory of neutral
kaons is that in this case such a product should be real.
\end{abstract}

\section{Introduction}

Almost all properties of the neutral meson complex are  described
by solving the Schr\"{o}dinger--like evolution equation
\cite{LOY1}
--- \cite{data} (we use $\hbar = c = 1$ units)
\begin{equation}
i \frac{\partial}{\partial t} |\psi ; t \rangle_{\parallel} =
H_{\parallel} |\psi ; t \rangle_{\parallel}, \; \; \; (t \geq
t_{0}), \label{Schr-like}
\end{equation}
(where $t_{0}$ is the initial instant) for $|\psi ; t
\rangle_{\parallel}$ belonging to the subspace ${\cal
H}_{\parallel} \subset {\cal H}$ (where ${\cal H}$ is the state
space of the physical system under investigation), e.g., spanned
by orthonormal neutral  kaons states $|K_{0}\rangle, \;
|{\overline{K}}_{0}\rangle$, and so on, (then states corresponding
to  the decay products belong to ${\cal H} \ominus {\cal
H}_{\parallel} \stackrel{\rm def}{=} {\cal H}_{\perp}$), and
nonhermitian effective Hamiltonian $H_{\parallel}$ obtained
usually by means of  the Lee--Oehme--Yang (LOY) approach
\cite{LOY1} --- \cite{data} (within the use of the
Weisskopf--Wigner approximation (WW) \cite{ww}):
\begin{equation}
H_{\parallel} \equiv M - \frac{i}{2} \, \Gamma, \label{H-eff}
\end{equation}
where $M = M^{+}, \; \Gamma = {\Gamma}^{+}$ are $(2 \times 2)$
matrices. In a general case $H_{||}$ can depend on time $t$,
$H_{||} \equiv H_{||}(t)$, \cite{horwitz,acta-1983}.

Usually, solutions of the evolution equation (\ref{Schr-like}) are
expressed in terms of the  eigenvectors of $H_{||}$. Generally, in
the case of two-dimensional subspace ${\cal H}_{||}$ the
eigenvectors of $H_{\parallel}$ acting in this ${\cal H}_{||}$
will be denoted as $|l\rangle , |s\rangle  $. In the general case
solutions of the eigenvalue problem for $H_{||}$
\begin{equation}
H_{||}\;|l(s)\rangle = \mu_{l(s)} \;|l(s)\rangle , \label{H||l>}
\end{equation}
have the following form \cite{ijmpa-1992,ijmpa-1993}
\begin{equation}
|l(s)\rangle    =  N_{l(s)} \Big(|{\bf 1}\rangle  -
{\alpha}_{l(s)}|{\bf 2}\rangle  \Big) , \label{urb-ls}
\end{equation}
where $|{\bf 1}\rangle$ stands for  the  vectors of the
$|K_{0}\rangle, \; |B_{0}\rangle$ type and $|{\bf 2}\rangle$
denotes antiparticles of  the particle "1":
$|{\overline{K}}_{0}\rangle, \; {\overline{B}}_{0}\rangle$,
$\langle{\bf j}|{\bf k}\rangle = {\delta}_{jk}$, $(j,k =1,2)$,
\begin{equation}
N_{l(s)} = \frac{1}{\sqrt{1 + |\alpha_{l(s)}|^{2}}}
=N_{l(s)}^{\ast}, \label{urb-N-ls}
\end{equation}
and
\begin{equation}
{\alpha}_{l(s)} = \frac{h_{z} - (+) h}{h_{12}},
\label{urb-alpha-ls}
\end{equation}
\begin{equation}
{\mu}_{l(s)} = h_{0} +(-) h \equiv m_{l(s)} - \frac{i}{2}
{\gamma}_{l(s)}. \label{urb-mu-ls}
\end{equation}
Quantities $m_{l(s)}, {\gamma}_{l(s)}$ are real, and
\begin{eqnarray} h_{0} & = & \frac{1}{2}(h_{11} + h_{22}),
\label{urb-h-0} \\ h &
\equiv  &  \sqrt{ h_{z}^{2} + h_{12} h_{21} }, \label{urb-h-a} \\
h_{z} & = &  \frac{1}{2} (h_{11}  -  h_{22}),   \label{urb-h-z}
\end{eqnarray}
\begin{equation}
h_{jk}  =  \langle{\bf j}|H_{\parallel}|{\bf k}\rangle, \;
(j,k=1,2). \label{urb-h-jk}
\end{equation}
In the case of neutral kaons  eigenvectors  of $H_{||}$ are
identified with the long, $|K_{L}\rangle$, (vector $|l\rangle $)
and short, $|K_{S}\rangle$, (vector $|s\rangle $) living
superpositions of $K_{0}$ and $\overline{K_{0}}$. This
identification of vectors $|l(s)\rangle$ with states
$|K_{L(S)}\rangle$ corresponds to the standard phase convention
for CP transformation: ${\cal C}{\cal P} |{\bf  1}> =  - |{\bf
2}>$,  ${\cal C}{\cal P} | {\bf 2}> = - |{\bf 1}>$. Within this
phase convention for systems preserving CP symmetry one has
$K_{L(S)} \rangle \leftrightarrow |K_{2(1)} \rangle$, where
vectors $|K_{1(2)} \rangle$:
\begin{equation}  |K_{1(2)}> \stackrel{\rm def}{=}  2^{-1/2}
(|{\bf  1}>  -  (+)  |{\bf  2}>) , \label{K1-K2}
\end{equation}   are the  eigenvectors of  the  $\cal{CP}$--operator
for the eigenvalues $\;+ (-1)$.

The  following identities are true for ${\mu}_{l}$ and
${\mu}_{s}$,
\begin{eqnarray}
{\mu}_{l}  +  {\mu}_{s}  &  =  & h_{11} + h_{22}  \equiv {\rm
Tr}\, H_{||}, \label{urb-mu-l+mu-s}
\\ {\mu}_{l} - {\mu}_{s} &  =  &  2h \stackrel{\rm def}{=}  \Delta
\mu = \Delta m - \frac{i}{2} \Delta \gamma \label{urb-2h=mu-mu}  \\
{\mu}_{l}  \, {\mu}_{s}  &  =  &  h_{11} h_{22} -   h_{12}h_{21}
\equiv \det\, H_{||} , \label{urb-mu-l-mu-s}
\end{eqnarray}
where
\begin{equation}
\Delta m = m_{l} - m_{s} = (\Delta m)^{\ast}, \;\;\;\;\;\; \Delta
\gamma = \gamma_{l} - \gamma_{s} = (\Delta \gamma )^{\ast}.
\label{dm}
\end{equation}

Sometimes it is convenient to express vectors $|l\rangle$ and
$|s\rangle$  as  follows \cite{cronin-1981} -- \cite{comins}
\begin{equation} |l(s)\rangle \equiv
\frac{1}{ \sqrt{ 2(1  +  |{\varepsilon}_{l(s)}|^{2})}}\, [ (1 +
\varepsilon_{l(s)})|{\bf 1} \rangle + ( -1)(1 -
{\varepsilon}_{l(s)}) |{\bf 2} \rangle ] . \label{Kl-Ks}
\end{equation}
This form of $|l\rangle$ and $|s\rangle$ is used in many papers
when possible departures from CP-- or CPT--symmetry in the system
considered are discussed. The following  parameters are used to
describe the scale  of CP-- and possible  CPT  -- violation
effects \cite{cronin-1981} -- \cite{comins}
\begin{equation} \varepsilon
\stackrel{\rm def}{=} \frac{1}{2} ( {\varepsilon}_{s}  +
{\varepsilon}_{l}  ) , \label{epsilon}
\end{equation}  \begin{equation}  \delta \stackrel{\rm def}{=}
\frac{1}{2} (  {\varepsilon}_{s}  - {\varepsilon}_{l}  )  .
\label{delta} \end{equation} According  to the   standard meaning,
$\varepsilon$ describes violations of CP--symmetry and $\delta$ is
considered as  a  CPT--violating parameter \cite{LOY2} --
\cite{dafne}. Such an interpretation of  these parameters follows
from properties of LOY theory of  time evolution  in  the subspace
of neutral kaons \cite{LOY1,LOY2}. We have
\begin{eqnarray} \varepsilon & = & \frac{h_{12}  -
h_{21}}{D} \label{epsilon-hjk} \\ \delta & = & \frac{h_{22}  -
h_{11} }{D},  \label{delta-hjk}
\end{eqnarray} where
\begin{equation} D  \stackrel{\rm def}{=}  h_{12} + h_{21} +  \Delta
\mu . \label{D} \end{equation}

Using identities (\ref{urb-mu-l+mu-s}) -- (\ref{urb-mu-l-mu-s})
and relations expressing $\varepsilon_{l}, \varepsilon_{s}$ by
matrix elements $h_{jk}$ and $\mu_{l}, \mu_{s}$ one can find the
following equations
\begin{eqnarray}  h_{11}  -  h_{22}  &  =  &  \Delta  \mu
\frac{{\varepsilon}_{l} - {\varepsilon}_{s}}{ 1 -
{\varepsilon}_{l} {\varepsilon}_{s} } , \label{h11-h22} \\ h_{12}
+ h_{21} & = & \Delta \mu   \frac{1   +{\varepsilon}_{l}
{\varepsilon}_{s}   }{
1    - {\varepsilon}_{l} {\varepsilon}_{s} } , \label{h12+h21}  \\
h_{12}  - h_{21}   & = & \Delta    \mu    \frac{{\varepsilon}_{l}
+ {\varepsilon}_{s}}{ 1  -  {\varepsilon}_{l}  {\varepsilon}_{s} }
, \label{h12-h21}
\end{eqnarray} These equations   are    valid for arbitrary values
of ${\varepsilon}_{l(s)}$ and for the exact $H_{||}$ as well as
for any approximate $H_{||}$.

Experimentally measured  values  of  parameters
${\varepsilon}_{l}, {\varepsilon}_{s}$  are  very  small  for
neutral  kaons.  Assuming
\begin{equation}    |{\varepsilon}_{l}|     \ll     1,     \;     \;
|{\varepsilon}_{s}|  \ll  1,   \label{e-ls<<1}
\end{equation}
from ({\ref{h11-h22}) one finds:
\begin{equation} h_{11}  -  h_{22}
\simeq ({\mu}_{l} - {\mu}_{s}) ({\varepsilon}_{l}  -
{\varepsilon}_{s}  ), \label{h11-h22-app}
\end{equation}    and
({\ref{h12+h21})     implies
\begin{equation} h_{12}  +  h_{21}  \simeq  {\mu}_{l}  -  {\mu}_{s},
\label{h12+h21-app}     \end{equation}
and (\ref{h12-h21}) gives
\begin{equation} h_{12} -  h_{21}  \simeq  ({\mu}_{l}  -  {\mu}_{l})
({\varepsilon}_{l}     +      {\varepsilon}_{s}).
\label{h12-h21-app}
\end{equation}
Relation (\ref{h11-h22-app}) means that in  the  considered case
of small values   of   parameters $|{\varepsilon}_{l}|,
|{\varepsilon}_{s}|$, the quantity  $D$ (\ref{D}) appearing in
formulae for $\delta$ and $\varepsilon$ approximately equals
\begin{equation} D \simeq 2({\mu}_{l} - {\mu}_{s})  \equiv 2
\Delta \mu .  \label{D-app}
\end{equation}

In the standard approach to the description of properties of the
neutral kaon complex many relations connecting parameters
characterizing neutral kaons follow from the properties of the
scalar product of state vectors $|K_{S}\rangle, |K_{L}\rangle$.
The aim of this paper is to analyze in the general case detailed
properties of the scalar product of eigenvectors $|l\rangle$ and
$|s\rangle$ depending on CP and CPT transformations properties of
the total system under considerations.

\section{General properties of the product $\langle s|l\rangle$}

Let us analyze the product $\langle s|l \rangle$ in the case of a
general $H_{||}$ without any assumptions about CP-- or
CPT--symmetries of the system under considerations. From
(\ref{urb-ls}) one finds
\begin{equation}
\langle s|l\rangle = N_{s}\,N_{l}\, (1 +
\alpha_{s}^{\ast}\,\alpha_{l} \,). \label{<s|l>}
\end{equation}
The important question is whether the product $\langle s|l
\rangle$ is real, $\langle s|l \rangle \equiv (\langle s|l
\rangle)^{\ast}$, or not, $\langle s|l \rangle \neq (\langle s|l
\rangle)^{\ast}$. It is obvious that the answer to this question
depends on the  properties of the product
$\alpha_{s}^{\ast}\,\alpha_{l}$. From (\ref{urb-alpha-ls}) it
follows that
\begin{equation}
\alpha_{s}^{\ast}\,\alpha_{l} = \frac{1}{|h_{12}|^{2}} \Big[
\Big(|h_{z}|^{2} - |h|^{2}\Big) + 2i\, \Im \,(h_{z}\,h^{\ast})
\Big], \label{a_sxa_l}
\end{equation}
where $\Im\,(z)$ denotes the imaginary part of the complex number
$z$ $\;$($\Re\, (z)$ is the real part of $z$). So the trivial
conclusion is that
\begin{equation}
\langle s|l \rangle = (\langle s|l \rangle)^{\ast} \equiv \langle
l|s\rangle \;\;\;\Leftrightarrow \;\;\; \Im\, (h_{z}\,h^{\ast}) =
0. \label{<->=0}
\end{equation}
Taking into account the identity (\ref{urb-2h=mu-mu}) one has $h =
\frac{1}{2} \Big( \Delta m - \frac{i}{2} \Delta \gamma \Big)$ and
thus it can be easy found that
\begin{equation}
\Im\, (h_{z}\,h^{\ast}) \;=\; \frac{1}{2} \,(\Im \,h_{z})\;(\Delta
m) \,+ \,\frac{1}{4}\, (\Re \, h_{z})\;(\Delta \gamma).
\label{Im-hz-h}
\end{equation}

From this relation it is seen that if $h_{z} = 0$, that is, if
$(h_{11} - h_{22}) = 0$ (see (\ref{urb-h-z})) then $\Im\,
(h_{z}\,h^{\ast})\equiv 0$. This result does not depend on the
values of $\Delta m$ and $\Delta \gamma$. So, if $h_{z}=0$ then
the scalar product $\langle s|l \rangle$ must be real.

Now let us suppose  that $h_{z} \neq 0$. In order to draw some
conclusions  about $\Im\, (h_{z}h^{\ast})$ in this case one should
rewrite $\Delta \mu, \; \Delta m, \Delta \gamma$ and $(h_{11} -
h_{22})$ in a  more convenient form. If the superweak phase
$\phi_{SW}$ \cite{dafne,data} is used,
\begin{equation}
\tan\,\phi_{SW} = \frac{2(m_{l} - m_{s})}{\gamma_{s} - \gamma_{l}}
\equiv - \,\frac{2\,\Delta \mu}{\Delta \gamma}, \label{phi_sw}
\end{equation}
then one can find that
\begin{equation}
\Delta m = - |\Delta \mu | \sin\,\phi_{SW}, \;\;\;\;\;\;
\frac{\Delta \gamma}{2} = |\Delta \mu | \cos\, \phi_{SW}.
\label{dm-phi}
\end{equation}
Next one should find a similar expression for $\Re\, h_{z}$ and
$\Im\,h_{z}$. One has
\begin{equation}
h_{jj} = \Re \, (h_{jj})\; + i\,\Im\,(h_{jj}),\label{h-jj}
\end{equation}
$(j=1,2)$, where
\begin{equation}
\Re \, (h_{jj} )\, \equiv \,M_{jj}, \;
\;\;\;\;\;\Im\,(h_{jj})\,\equiv \, - \frac{1}{2}{\Gamma}_{jj}.
\label{M+G-jj}
\end{equation}
Using the following definitions
\begin{equation}
\Delta M = M_{11} - M_{22}, \;\;\;\; \Delta \Gamma = \Gamma_{11} -
\Gamma_{22}, \label{DM}
\end{equation}
one can write that
\begin{equation}
h_{11}- h_{22} = \Delta M - \frac{i}{2}\Delta \Gamma  \equiv 2
h_{z}. \label{h11-h22=DM}
\end{equation}
Next, if another phase $\phi_{z}$ is introduced analogously to the
superweak phase $\phi_{SW}$ by means of the relation
\begin{equation}
\tan\,\phi_{z} = - \,\frac{2\,\Delta M}{\Delta \Gamma},
\label{phi-z}
\end{equation}
then one finds that
\begin{equation}
\Re\,h_{z} = - |h_{z}|\, \sin\,\phi_{z},\;\;\;\; \Im\,h_{z} =
-|h_{z}|\,\cos\,\phi_{z}. \label{Re-h-z}
\end{equation}
Thus, using (\ref{dm-phi}) and (\ref{Re-h-z}) relation
(\ref{Im-hz-h}) can be rewritten in a compact and convenient form
\begin{equation}
\Im\,(h_{z}\,h^{\ast}) = \frac{1}{2}\,|\Delta \mu |\,|h_{z}|\,
\sin\,(\phi_{SW} - \phi_{z}). \label{Im-h-z-final}
\end{equation}

Note that, e.g.., if $\Delta M \neq 0$ and $\Delta \Gamma = 0$ then
$\phi_{z} = \frac{1}{2}\pi + n\pi$, $(n = 0,\pm 1,\pm 2, \ldots)$.
On the other hand there is $\phi_{SW} \approx43,5^{0}$  in the
case of neutral $K$ mesons (see \cite{branco,dafne,data}).

So, for such values of $\Delta m$ and $\Delta \gamma$ (and thus
$\phi_{sw}$) which are characteristic of neutral mesons
(neutral kaons, neutral $B$ mesons subsystems, etc. \cite{data})
the condition $h_{z} \neq 0$ causes that $\Im\, (h_{z}\,h^{\ast})
\neq 0$ and thus by (\ref{<s|l>}) and (\ref{a_sxa_l}) that
$\langle s|l\rangle \neq \langle l|s\rangle \equiv (\langle s|l
\rangle )^{\ast}$.

All the above analysis leads to the conclusion that in the case on
neutral mesons the following theorem holds\\
\hfill\\
{\bf Theorem}

For the values of $\Delta m$ and $\Delta \gamma$ which are typical
for neutral meson complexes
\begin{equation}
\langle s|l \rangle = (\langle s|l \rangle)^{\ast} \equiv \langle
l|s\rangle \;\;\Leftrightarrow \;\; (h_{11} - h_{22}) = 0.
\label{h11=h22-s|l}
\end{equation}
\hfill\\

It is interesting to confront this observation with the properties
of matrix elements, $h_{jk}$, of the approximate as well as the
exact effective Hamiltonians for neutral meson complex following
from the CP-- or CPT--symmetries of the total system under
considerations. The standard approach to the description of
properties of neutral mesons is based on the LOY effective
Hamiltonian, $H_{LOY}$. Taking $H_{||} = H_{LOY}$ and assuming
that the CPT invariance holds in the system considered one easily
finds the standard result of the LOY approach
\begin{equation}
h_{11}^{LOY}  = h_{22}^{LOY}, \label{LOY-h=h}
\end{equation}
where $h_{jk}^{LOY} = \langle {\bf j}|H_{LOY}|{\bf k}\rangle,\;\;
(j,k =1,2)$. Therefore within the LOY theory the property that in
a CPT invariant system $\langle K_{S}|K_{L}\rangle = (\langle
K_{S}|K_{L}\rangle )^{\ast} \equiv \langle K_{L}|K_{S}\rangle$, is
considered as quite obvious and unquestionable. This is one of the
standard results of the LOY theory of neutral meson complexes. The
question is whether such a property of the scalar product under
considerations  holds in the case of the exact effective
Hamiltonian for the neutral mesons complex or not.\

\section{CP and CPT transformations \\ and
the exact $H_{||}$}

Solutions of the Schr\"{o}dinger--like equation (\ref{Schr-like})
can be written in matrix  form  and such a  matrix  defines  the
evolution operator (which    is usually nonunitary)
$U_{\parallel}(t)$ acting in ${\cal H}_{\parallel}$:
\begin{equation}
|\psi ; t \rangle_{\parallel}  \stackrel{\rm def}{=}
U_{\parallel}(t) |\psi \rangle_{\parallel}, \label{U||}
\end{equation}
where,
\begin{equation}
|\psi \rangle_{\parallel} \equiv q_{1}|{\bf 1}\rangle + q_{2}|{\bf
2}\rangle, \label{psi-t0}
\end{equation}
is the initial state of the system, $|{\psi}\rangle_{||}\, \equiv
|{\psi},t = t_{0}\rangle_{||}\, \in \, {\cal H}_{||}$. CP and CPT
transformation properties of the matrix elements, $h_{jk}$, of the
exact effective Hamiltonian, $H_{||}$, can be extracted from the
suitable properties of the exact evolution operator $U_{||}(t)$.
The exact evolution operator $U_{||}(t)$ has the following form
\cite{plb-2002}
\begin{equation}
U_{||}(t) = P U(t)P, \label{ku-U||}
\end{equation}
$P$ is the projection operator onto subspace ${\cal
H}_{\parallel}$ and $U(t)$ is the total unitary evolution
operator, which solves the Schr\"{o}\-din\-ger equation
\begin{equation}
i \frac{\partial}{\partial t} U(t)|{\psi} \rangle_{||} = H
U(t)|{\psi}\rangle_{||}, \; \; U(t = t_{0}) = I, \label{Schr}
\end{equation}
where $I$ is the unit operator in $\cal H$ and $H$ is the total
(selfadjoint) Hamiltonian acting in $\cal H$. In the considered
case the projector $P$ can be defined as follows
\cite{ijmpa-1993,plb-2002}
\begin{equation}
P  =  |{\bf 1} \rangle  \langle {\bf 1}| + |{\bf 2}\rangle \langle
{\bf 2}|.  \label{urb-P}
\end{equation}
One has
\begin{equation}
{\cal H}_{||} = P {\cal H}, \;\;\;\;\; {\cal H}_{\perp} = (I -
P){\cal H} \stackrel{\rm def}{=} Q {\cal H} . \label{urb-P+Q}
\end{equation}
The evolution operator $U_{||}(t)$ has a  nontrivial form only if
\begin{equation}
[P, H] \neq 0, \label{ku-[P,H]}
\end{equation}
and only then transitions of states from ${\cal H}_{||}$ into
${\cal H}_{\perp}$ and vice versa, i.e., decay and regeneration
processes, are allowed.

Within the matrix representation one can write \cite{plb-2002}
\begin{equation}
U_{||}(t) \equiv \left(
\begin{array}{cc}
{\rm \bf A}(t) & {\rm \bf 0} \\
{\rm \bf 0} & {\rm \bf 0}
\end{array} \right),
\label{A(t)}
\end{equation}
where ${\rm \bf 0}$ denotes the suitable zero submatrices and a
submatrix ${\rm \bf A}(t)$ is the $(2 \times 2)$ matrix acting in
${\cal H}_{||}$,
\begin{equation}
{\rm \bf A}(t) = \left(
\begin{array}{cc}
A_{11}(t) & A_{12}(t) \\
A_{21}(t) & A_{22}(t)
\end{array} \right), \label{A(t)=}
\end{equation}
and $A_{jk}(t) =  \langle {\bf j}|U_{||}(t)|{\bf k}\rangle \equiv
\langle {\bf j}|U(t)|{\bf k}\rangle$, $(j,k =1,2)$.

Now if we assume that
\begin{equation}
[ \Theta, H] = 0, \label{urb-[CPT,H]}
\end{equation}
(where ${\Theta} = {\cal C}{\cal P}{\cal  T}$ is an antiunitary
operator with unitary ${\cal C}, \; {\cal P}$ and antiunitary
${\cal T}$ denoting operators realizing charge conjugation, parity
and time reversal for vectors in  $\cal H$ respectively), then one
easily finds that \cite{gibson}, \cite{plb-2002} ---
\cite{nowakowski-1999}
\begin{equation}
A_{11}(t) = A_{22}(t). \label{A11=A22}
\end{equation}
The assumption (\ref{urb-[CPT,H]}) gives no relations between
$A_{12}(t)$ and $A_{21}(t)$.

If the system under considerations is assumed to be CP invariant,
\begin{equation}
[{\cal CP},H]=0, \label{[CP,H]}
\end{equation}
then using the following, most general, phase convention
\begin{equation}
{\cal CP}|{\bf 1}\rangle = e^{-i\alpha}|{\bf 2}\rangle, \;\;\;
{\cal CP}|{\bf 2}\rangle = e^{+i\alpha}|{\bf 1}\rangle,
\label{CP1=2}
\end{equation}
(instead of the standard one: ${\cal C}{\cal  P}  |{\bf 1}> = -
|{\bf 2}>$, ${\cal C}{\cal P} | {\bf 2}> = - |{\bf 1}>$) one
easily finds that for the diagonal matrix elements of the matrix
${\bf A(t)}$ the relation (\ref{A11=A22}) holds in this case also,
and that for the off--diagonal matrix elements
\begin{equation}
A_{12}(t) = e^{2i \alpha}A_{21}(t). \label{A12=A21}
\end{equation}

This means that if the CP symmetry is conserved in the system
containing the subsystem of neutral mesons, then for every $t>0$
there must be
\begin{equation}
\mid \frac{A_{12}(t)}{A_{21}(t)}\mid \;= 1\; \equiv\; {\rm const.}
\label{A12/A21=1}
\end{equation}

Now let us consider the case when CP symmetry is violated,
\begin{equation}
[{\cal CP},H] \neq 0. \label{[CP,H]no-0}
\end{equation}
For our considerations it is convenient to decompose the total
Hamiltonian $H$ into two parts \cite{Lee-qft,Bigi},
\begin{equation}
H \equiv H_{+} + H_{-}, \label{H=H+H-}
\end{equation}
where
\begin{equation}
H_{\pm} = \frac{1}{2} [H \pm ({\cal CP})H({\cal CP})^{+}].
\label{H-pm}
\end{equation}
Under $\cal CP$, $H_{+}$ is even and $H_{-}$ is odd,
\begin{equation}
({\cal CP}) H_{\pm}({\cal CP})^{+} \,= \, \pm H_{\pm}.
\label{CP-H-pm}
\end{equation}
(Note that if relation (\ref{[CP,H]}) holds then $H_{-} \equiv
0$). Now using relations (\ref{H=H+H-}) --- (\ref{CP-H-pm}) one
can easy conclude that
\begin{equation}
({\cal CP})H({\cal CP})^{+} \; \equiv \; H \, - \,2H_{-}.
\label{CP-H-CP+}
\end{equation}
This result helps one to solve the problem of how  the solution,
$U(t)$, to the Schr\"{o}dniger equation (\ref{Schr}) transforms
under $\cal CP$. So let us define $U_{CP}(t) \stackrel{\rm def}{=}
({\cal CP}) U(t)({\cal CP})^{+}$, where $U(t)$ solves
Eq.(\ref{Schr}). Starting from Eq.(\ref{Schr}) one obtains
\begin{eqnarray}
i \frac{\partial}{\partial t} U_{CP}(t) &=& (H \,-\,
2H_{-})U_{CP}(t) \label{U-CP-1}\\
&\equiv & HU_{CP}(t) \,- \, 2H_{-}U_{CP}(t), \label{U-CP-2}
\end{eqnarray}
with the initial condition  $U_{CP}(0) = I$. The solution, $U(t)$,
of Eq.(\ref{Schr}) is the "free" solution for Eq.(\ref{U-CP-2})
and thus the solution of this last equation can be expressed as
follows \cite{kato}
\begin{equation}
U_{CP}(t) \; = \; U(t)\, + \, 2i\int_{0}^{t} U(t - \tau
)\,H_{-}\,U_{CP}(\tau )\,d\tau .\label{U-CP-sol}
\end{equation}
Of course from (\ref{U-CP-1}) it follows that $U_{CP}(t) = \exp
[-it(H - 2H_{-})t]$ but this formula is much less convenient than
(\ref{U-CP-sol}).

Assuming that the system under consideration is not CP invariant
and using (\ref{CP1=2}) it is easy to find that
\begin{equation}
A_{12}(t) \equiv  e^{2i \alpha} \langle {\bf 2}|U_{CP}(t)|{\bf 1}
\rangle . \label{A12-A21-CP}
\end{equation}
Next, inserting there $U_{CP}(t)$ given by (\ref{U-CP-sol}) yields
\begin{equation}
A_{12}(t) \,=\,  e^{2i \alpha}A_{21}(t) \, +\,2i e^{2i \alpha}
\langle {\bf 2}| \int_{0}^{t} U(t - \tau )\,H_{-}\,U_{CP}(\tau
)\,d\tau |{\bf 1}\rangle.
\end{equation}
From this last relation one infers that when CP symmetry is
violated then for $t
> 0$ there must be
\begin{equation}
\mid \frac{A_{12}(t)}{A_{21}(t)} {\mid}^{\,2} \,\equiv \, 1 + \,
|r_{21}(t)|^{2}\,+\, 2 \,\Re\,(r_{21}(t)),
\;\;\;(t>0),\label{A12/A21-CP-viol}
\end{equation}
where
\begin{equation}
r_{21}(t) \;=\; \frac{2i}{A_{21}(t)}\, \int_{0}^{t} \langle {\bf
2}|U(t - \tau )\,H_{-}\,U_{CP}(\tau )|{\bf 1}\rangle\,d\tau  ,
\label{r21}
\end{equation}
and $r_{21}(t) \neq 0$ for $t>0$.

Let us analyze the simplest case when $t$ is very short, $t
\approx 0$, but still $t > 0$ and $\langle {\bf 2}|H|{\bf
1}\rangle \neq 0$. Then after some algebra one finds
\begin{equation}
{\vrule \, \frac{A_{12}(t)}{A_{21}(t)} \, \vrule }^{\,2}
\begin{array}[t]{l} \vline \, \\ \vline
\, {\scriptstyle \,0<t \approx 0 }
\end{array}
\simeq \, 1 \, + \,
4 \,{\vrule \, \frac{\langle {\bf 2}|H_{-}|{\bf 1}\rangle }{\langle {\bf
2}|H|{\bf 1}\rangle}\, \vrule}^{\,2}
 \, - \, 4 \, \Re\, \Big( \frac{\langle
{\bf 2}|H_{-}|{\bf 1}\rangle }{\langle {\bf 2}|H|{\bf 1}\rangle}
\Big) \; \neq \; 1.\label{A12/A21-t=0}
\end{equation}
If $\langle {\bf 2}|H|{\bf
1}\rangle = 0$ but $\langle {\bf 2}|H^{2}|{\bf
1}\rangle \neq 0$ then for very short $t$ one has
\begin{eqnarray}
{\vrule \, \frac{A_{12}(t)}{A_{21}(t)} \, \vrule }^{\,2}
\begin{array}[t]{l} \vline \, \\ \vline
\, {\scriptstyle \,0<t \approx 0 }
\end{array}
& \simeq & \, 1 \, + \,
4\, { \vrule \,\frac{\langle {\bf 2}|(H_{+}H_{-} + H_{-}H_{+})|{\bf 1}\rangle }{\langle {\bf
2}|H^{2}|{\bf 1}\rangle}\,\vrule \,}^{\,2} \nonumber \\
 & & \;\;\; - \, 4 \, \Re\, \Big( \frac{\langle
{\bf 2}|(H_{+}H_{-} + H_{-}H_{+})|
{\bf 1}\rangle }{\langle {\bf 2}|H^{2}|{\bf 1}\rangle}
\Big) \; \neq \; 1.\label{A12/A21-t=0+H12no0}
\end{eqnarray}

Note that these results and  (\ref{A12/A21-CP-viol}) are the quite
general and that they do not depend on any model or
approximation used. Relations (\ref{A12/A21-CP-viol}),
(\ref{A12/A21-t=0}) and (\ref{A12/A21-t=0+H12no0})
prove that if the property (\ref{[CP,H]no-0})
holds in the system, that is if the CP symmetry is violated, then
in a such system the modulus of the ratio
$\frac{A_{12}(t)}{A_{21}(t)}$ must be different from 1 for every $t>0$,
\begin{equation}
[{\cal CP}, H] \, \neq \,0 \;\;\; \Rightarrow \;\;\; \mid
\frac{A_{12}(t)}{A_{21}(t)} {\mid} \, \neq \, 1, \;\;\;\;(t > 0).
\label{A12/A21-neq-1}
\end{equation}
The importance of this result consists in the fact that it is the
rigorous consequence of only two assumptions. The first is that
the real properties of the system follow from the solutions of the
Schr\"{o}dinger Equation (\ref{Schr}). The second one is that the
total selfadjoint Hamiltonian $H$ does not commute with the $\cal
CP$ operator. Apart from these two assumptions no additional model
assumptions or approximations were used in order to prove
(\ref{A12/A21-neq-1}). In particular, no properties of the
eigenvectors $|l\rangle , |s\rangle$ for the effective Hamiltonian
$H_{||}$ and no assumptions about their form were used in the
above considerations leading to the conclusion
(\ref{A12/A21-neq-1}).

So, we already have all the necessary CP-- and CPT--transformation
properties of the matrix elements of the exact evolution operator
$U_{||}(t)$ for the subspace of neutral mesons,${\cal H}_{||}$,
and now we can extract from them the suitable properties of the
matrix elements of the exact effective Hamiltonian for this
subspace. One can find the necessary properties of the matrix
elements of $H_{||}$ by analyzing the following identity
\cite{horwitz,acta-1983,plb-2002,bull,pra}
\begin{equation}
H_{||} \equiv H_{||}(t) = i \frac{\partial U_{||}(t)}{\partial t}
[U_{||}(t)]^{-1}, \label{H||2a}
\end{equation}
where $[U_{||}(t)]^{-1}$ is defined as follows
\begin{equation}
U_{||}(t) \, [U_{||}(t)]^{-1} = [U_{||}(t)]^{-1} \, U_{||}(t) \, =
\, P. \label{U^-1}
\end{equation}
(Note that the identity (\ref{H||2a}) holds, independent of
whether $[P,H] \neq 0$ or $[P,H]=0$). The expression (\ref{H||2a})
can be rewritten using the matrix ${\bf A}(t)$

\begin{equation}
H_{||}(t) \equiv  i \frac{\partial {\bf A}(t)}{\partial t} [{\bf
A}(t)]^{-1}. \label{H||2b}
\end{equation}
Relations (\ref{H||2a}), (\ref{H||2b}) must be fulfilled by the
exact as well as by every approximate effective Hamiltonian
governing the time evolution in every two dimensional subspace
${\cal H}_{||}$ of states $\cal H$
\cite{horwitz,ijmpa-1992,bull,pra}.

It is easy to find from (\ref{H||2b}) the general formulae for the
diagonal matrix elements, $h_{jj}$, of $H_{||}(t)$, in which we
are interested. We have \cite{plb-2002}
\begin{eqnarray}
h_{11}(t) &=& \frac{i}{\det {\bf A}(t)} \Big( \frac{\partial
A_{11}(t)}{\partial t} A_{22}(t) - \frac{\partial
A_{12}(t)}{\partial t} A_{21}(t) \Big), \label{h11=} \\
h_{22}(t) & = & \frac{i}{\det {\bf A}(t)} \Big( - \frac{\partial
A_{21}(t)}{\partial t} A_{12}(t) + \frac{\partial
A_{22}(t)}{\partial t} A_{11}(t) \Big). \label{h22=}
\end{eqnarray}
Using (\ref{h11=}), (\ref{h11=}) the difference $(h_{11} - h_{22})
= 2h_{z}$, whose properties are crucial  for the question whether
the product $\langle s|l\rangle$ is real or not, can be expressed
as follows \cite{plb-2002}
\begin{eqnarray}
h_{11}(t) - h_{22}(t) &=& i \frac{1}{\det {\bf A}(t)}\, \Big\{
A_{11}(t) \, A_{22}(t)  \; \frac{\partial}{\partial t} \ln
\Big(\frac{A_{11}(t)}{A_{22}(t)} \Big) \nonumber \\
&& \;\;\;\;\;\;\;\;\;\;\;\;\; -\, A_{12}(t) \, A_{21}(t)  \;
\frac{\partial}{\partial t} \ln \Big(\frac{A_{12}(t)}{A_{21}(t)}
\Big) \Big\}. \label{h11-h22=1}
\end{eqnarray}

At this point one should use the fact that an important relation
between amplitudes $A_{12}(t)$ and $A_{21}(t)$ is described by the
famous Khalfin's Theorem \cite{chiu}
--- \cite{leonid-fp}, \cite{kabir,kabir+mitra}. This Theorem states that
in the case of unstable states, if amplitudes $A_{12}(t)$ and
$A_{21}(t)$ have the same time dependence
\begin{equation}
r(t) \stackrel{\rm def}{=} \frac{A_{12}(t)}{A_{21}(t)} = {\rm
const} \equiv r, \label{r=const},
\end{equation}
then there must be $|r| = 1$.

The proof of this theorem is rigorous and it does not use the CP--
or CPT-- transformation properties of the system considered.

Now one is ready to examine consequences of the assumptions that
$(h_{11}(t) - h_{22}(t)) = 0$ is admissible for $t >0$.  In such a
case an analysis of the expression (\ref{h11-h22=1}), relations
(\ref{A11=A22}), (\ref{A12/A21=1}), (\ref{A12/A21-CP-viol}),
(\ref{A12/A21-neq-1}) and the Khalfin's Theorem (\ref{r=const})
allows one to conclude that\\
\hfill\\
\hfill\\
\textbf{Conclusion 1.}\\
If $(h_{11}(t) - h_{22}(t)) = 0$ for $t>0$ then there must be\\
a)
\[ \frac{A_{11}(t)}{A_{22}(t)} = {\rm const.},\;\; {\rm and} \;\;
\frac{A_{12}(t)}{A_{21}(t)} = {\rm const.},\;\;
  ({\rm for}\;\; t > 0) ,\]
or,\\
b)
\[\frac{ A_{11}(t)}{A_{22}(t)} \neq {\rm const.},\;\; {\rm and} \;\;
\frac{A_{12}(t)}{A_{21}(t)} \neq {\rm const.},\;\;
  ({\rm for}\;\; t > 0) .\]
\hfill\\

The following interpretation of a) and b) follows from
(\ref{A11=A22}), (\ref{A12/A21=1}), (\ref{A12/A21-CP-viol}),
(\ref{A12/A21-neq-1}) and from the Khalfin's Theorem
(\ref{r=const}). Case a) means that CP--symmetry is conserved and
there is no information about CPT invariance. Case b) denotes that
the system under considerations is neither CP--invariant nor
CPT--invariant.

In our discussion the  CPT Theorem \cite{pauli}
--- \cite{wightman} can not be neglected. The CPT Theorem is
a fundamental theorem of axiomatic quantum field theory. It
follows from locality, Lorentz invariance and unitarity.  One
should also take into account another fact that there is no an
experimental evidence that CPT symmetry is violated
\cite{data,cronin-1964}. Therefore, the assumption that any
quantum theory of elementary particles should be CPT invariant
seems to be obvious. So let us assume that CPT symmetry is the
exact symmetry of the system under considerations, that is that
the condition (\ref{urb-[CPT,H]}) holds. In such a case the
relation (\ref{A11=A22}) holds. The consequence of this is that
the expression (\ref{h11-h22=1}) becomes simpler and it is easy to
prove that the following property must hold \cite{plb-2002}
\begin{equation}
h_{11}(t) - h_{22}(t) = 0 \; \; \Leftrightarrow \; \;
\frac{A_{12}(t)}{A_{21}(t)}\;\; = \; \; {\rm const.}, \; \; (t >
0). \label{h11-h22=0<=>}
\end{equation}

Now let us go on to analyze the  conclusions following from the
Khalfin's Theorem. CP noninvariance requires that $|r| \neq 1$
(see (\ref{A12/A21=1}), (\ref{A12/A21-CP-viol}),
(\ref{A12/A21-neq-1}), and also \cite{LOY1}
--- \cite{data}, \cite{chiu} --- \cite{nowakowski-1999}). This
means that in such a case there must be $r = r(t) \neq {\rm
const.}$. So, if in the system considered the properties
(\ref{urb-[CPT,H]}) and (\ref{[CP,H]no-0}) hold then, as it
follows from (\ref{h11-h22=0<=>}), at $t > 0$ there  must be
$(h_{11}(t) - h_{22}(t)) \neq 0$ in this system \cite{plb-2002}.
Thus, keeping in mind results (\ref{Im-h-z-final}) and
(\ref{h11=h22-s|l}) one can
state that following conclusion must be true\\
\hfill\\
\textbf{Conclusion 2.}

If CPT symmetry is the real symmetry of the system containing
neutral meson subsystem and CP symmetry is violated in this system
(i.e., if (\ref{urb-[CPT,H]}) and (\ref{[CP,H]no-0}) hold) then
there must be
\begin{equation}
\langle s|l \rangle \neq (\langle s|l \rangle)^{\ast} \equiv
\langle l|s\rangle . \label{s|l-neq-l|s}
\end{equation}

\section{Discussion}

As it was mentioned, the  CPT Theorem follows from basic
principles of quantum theory. Simply it is the mathematical
consequence of basic assumptions of quantum theory. There is no
evidence that the basic principles of quantum mechanics are
violated. There is also no evidence of CPT violation. In contrast
to the lack of evidence of the CPT noninvariance, the CP violation
is an experimental fact \cite{cronin-1981,data,cronin-1964}. This
means (by (\ref{A12/A21-neq-1})) that in the real system there
must be $|\frac{A_{12}(t)}{A_{21}(t)}| \neq 1$ for $(t
> 0)$ and therefore, due to the Khalfin's theorem (\ref{r=const})
and the relation (\ref{h11-h22=0<=>}), there must be $(h_{11}(t) -
h_{22}(t)) \neq 0$ for $(t > 0)$ in real systems. Thus the real
property of the system containing neutral mesons is that it must
be $\langle s|l \rangle \neq (\langle s|l \rangle)^{\ast} \equiv
\langle l|s\rangle$ (rather than $\langle s|l \rangle = (\langle
s|l \rangle)^{\ast} \equiv \langle l|s\rangle $).  This means that
one of the standard results of the LOY theory that in a CPT
invariant system $\langle K_{S}|K_{L}\rangle = (\langle
K_{S}|K_{L}\rangle )^{\ast} \equiv \langle K_{L}|K_{S}\rangle$,
(where $|K_{S}\rangle, \,|K_{L}\rangle$ correspond to $|s\rangle,
|l\rangle$), is wrong.

On the other hand, the assumption that the inner product $\langle
K_{S}|K_{L}\rangle$ should be real (or equivalently that there
should be $\Im\,\langle K_{S}|K_{L}\rangle = 0$) when CPT symmetry
holds was considered in the literature as the the fundamental
property of CPT invariant system allowing one to derive many
relations connecting parameters characterizing neutral $K$ system
and some constrains on these parameters \cite{LOY2} ---
\cite{data}, \cite{bell} --- \cite{tsai}, etc.

The result (\ref{s|l-neq-l|s}) means that all relations and
constrains obtained in this way  need not reflect real properties
of systems under consideration.  Simply, they may lead to wrong
conclusions obfuscating the  real properties of the neutral meson
systems and thus our opinion about the interactions causing decay
process of theses mesons. So within the standard LOY theory of
neutral meson complexes one should be very careful interpreting
the results of experiments with neutral mesons and in such a case
one can never be sure that this interpretation corresponds to the
real properties of the system under investigations. These
reservations also concern relations derived within the use of the
Bell--Steinberger relation. Properties of the inner product
$\langle K_{S}|K_{L}\rangle$ are crucial  for the interpretation
of such relations \cite{bell,tsai}. It seems that while performing
an analysis of the results of such experiments, only relations
connecting the parameters characterizing neutral meson complexes
which do not depend on any approximations  and which follow
directly from the general principles of quantum theory should be
taken into account. The mentioned above Khalfin's Theorem is an
example of such relations.

Note that if we assume that real properties of the system are
described by the solutions of the Schr\"{o}dinger Equation
(\ref{Schr}) (with $H=H^{+}$) then there can be $\langle
s|l\rangle = \langle l|s\rangle \equiv (\langle s|l\rangle
)^{\ast}$ for $h_{z} \neq 0$ only if  $\phi_{z} = \phi_{SW}$, that
is if
\begin{equation}
\frac{2\,\Delta M}{\Delta \Gamma} \equiv -\,\frac{\Re\,(h_{11} -
h_{22})}{\Im \,(h_{11} - h_{22})} = \frac{2\,(\mu_{l} -
\mu_{s})}{\gamma_{l} - \gamma_{s}}. \label{DM/DG}
\end{equation}
The result of the Fridrichs--Lee model \cite{chiu} calculations
performed in \cite{ijmpa-1993} with the assumption that values
of parameters of this model correspond to the  parameters of neutral
$K$ complex is the following  \cite{plb-2002}
\begin{equation}
\Re \, (h_{11}^{FL} - h_{22}^{FL}) = \Delta M^{FL} \sim 1,7 \times
10^{-13} \Im\,(\langle {\bf 1}|H|{\bf 2}\rangle)  \neq 0,
\label{FL-1}
\end{equation}
and
\begin{equation}
 \Im \, (h_{11}^{FL} - h_{22}^{FL})\equiv -
\frac{1}{ 2} \,\Delta \Gamma^{FL} = 0. \label{FL-2}
\end{equation}
So, the relation (\ref{DM/DG}) does not take place in the case of
the Fridrichs--Lee model. The same conclusion one can draw
analyzing the  experimentally obtained values $\Delta M = m_{K_{0}} -
m_{\bar{K}_{0}}$ and $\Delta \mu = m_{K_{L}} - m_{K_{S}},\; \Delta
\gamma = \Gamma_{K_{L}} - \Gamma_{K_{S}}$ \cite{data}.

The consequence of the result (\ref{s|l-neq-l|s}) is that
\begin{equation}
\frac{\langle s|l \rangle}{|\langle s|l \rangle|} =
e^{\textstyle{-i \vartheta}}, \label{theta}
\end{equation}
where, $ \vartheta \neq 0, \pm \pi, \pm 2\pi, \pm 3\pi, \ldots ,$
and so on. The phase $\vartheta$ can be easily found. Indeed,
using (\ref{Kl-Ks}) and parameters (\ref{epsilon}), (\ref{delta})
the product $\langle s|l\rangle$ can be expressed as follows
\begin{equation}
\langle s|l\rangle \equiv 2N [\,\Re \,({\varepsilon}) -i \, \Im \,
(\delta )\,], \label{s-l-1}
\end{equation}
where $N = N^{\ast} = [(1 + |{\varepsilon}_{s}|^{2}) (1 +
|{\varepsilon}_{l}|^{2})]^{- 1/2}$. Thus
\begin{equation}
\tan\,\vartheta = \frac{\Im\,(\delta)}{\Re\, (\varepsilon)}.
\label{tan-theta}
\end{equation}
So there is a direct connection between the phase $\vartheta$ and
the parameters $\delta$ and $\varepsilon$.

Some approximate estimations of $\vartheta$ can be obtained using
values of $\Im\,(\delta)$ and $\Re\,(\varepsilon)$ extracted from
experiments with neutral kaons. So, inserting into
(\ref{tan-theta}) values of $\Im\,(\delta)$, $\Re\,(\varepsilon)$
which can be found in \cite{data}, yields $\;\vartheta \;\lesssim
\; 0^{o}\,51'\,30''$. On the other hand taking into account, e.g.
some values of these parameters obtained in \cite{tsai2}, leads to
the result $\;\vartheta\; \lesssim\;1^{o}\,33'\,45'' $. The
mentioned above Fridrichs--Lee model calculations give similar
estimations for $\vartheta$.

The result (\ref{s|l-neq-l|s}) is the consequence of  the property
(\ref{A12/A21-neq-1}) that for $t > 0$ modulus of the ratio
$\frac{A_{12}(t)}{A_{21}(t)}$ must be different from unity if CP
symmetry is violated. Within the LOY theory such a property
follows from the properties of matrix elements of the approximate
LOY effective Hamiltonian. On the other hand, taking into account
the conclusions derived in \cite{nowakowski-2002}, the property
(\ref{A12/A21-neq-1}) can not considered to be the obvious.

In \cite{nowakowski-2002} it was found that within the LOY theory
the modulus of the similar ratio can be equal to one for some models
of interactions.  Note that the conclusion (\ref{A12/A21-neq-1})
does not depend on any approximation. It also does not depend on any
model of interactions.  As it was mentioned it is the simple
implication of two general assumptions: that the total Hamiltonian,
$H$, does not commute with the $\cal CP$ operator and that the
Schr\"{o}dinger equation describes correctly the system we are
interested in. This means that this relation reflects real
properties of system with CP violated.

All evolution equations for neutral meson complex have the form of
Eq. (\ref{Schr-like}) (see \cite{LOY1} --- \cite{data},
\cite{fermilab,babar}). Solutions of this equation are used to
describe time evolution of neutral mesons and mixing processes.
Solving this equation one can find amplitudes $A_{jk}(t)$,
($j,k=1,2$).  An important property of the ratio
$\frac{A_{12}(t)}{A_{21}(t)} = r(t)$ follows from the Khalfin's
Theorem (\ref{r=const}). The main result of this paper (i.e., the
property (\ref{s|l-neq-l|s})) and the earlier result that there
must be $h_{11}(t) - h_{22}(t)) \neq 0$ for $t >0$ in CPT
invariant system \cite{plb-2002} is the consequence of this
Theorem. (Note that the property $(h_{11}(t) - h_{22}(t)) \neq 0$
means by (\ref{delta}) and (\ref{delta-hjk}) that there must be $
\varepsilon_{l} \neq \varepsilon_{s}$ when CPT symmetry holds and
CP is violated (see also \cite{novikov})). One may want to
confront the Khalfin's Theorem with the experimental results,
which give $|1 - |r(t)|\,| \sim 10^{-3}$ = const. with some
limited accuracy (see, eg., \cite{data,babar}). This does not mean
that the Khalfin's Theorem is wrong. Simply effects connected with
the Khalfin's Theorem are very tiny and they seem to be beyond the
accuracy of recent experiments (see also \cite{kabir+mitra}). In
the light of the detailed model analysis given in \cite{chiu} the
conclusion that for $t \gg t_{0}$, $|r_{max}(t) - r_{min}(t)|
\stackrel{\rm def}{=} \Delta r < 10^{-11}$, seems to be
acceptable. Within the LOY approximation physical states,
$|l\rangle, |s \rangle$, decays exponentially. In general there
are tiny corrections to the exponential decay laws at very short
and very long times \cite{leonid-57}. The amplitudes $A_{jk}(t)$
calculated within the LOY, that is in fact within the WW
approximation give the result $r(t) = r_{LOY}$ = const. These
amplitudes calculated more accurately contains non-exponential and
non-oscillatory tiny corrections (see \cite{chiu,nowakowski-1999}
leading to varying in time $r(t)$ with the spectrum of changes
limited by $\Delta r < 10^{-11}$. In other words there is
\begin{equation}
r(t) = r_{LOY} + d(t), \label{d-r}
\end{equation}
where $d(t)$ varies in time $t$ and $|\,d(t)| \leq \Delta r$ for
$t \gg t_{0}$. (Note that the Khalfin's Theorem does not require
$d(t)$ to be large.) These corrections seems to be irrelevant for
many parameters describing neutral meson complex but they, and
therefore the consequences of the Khalfin's Theorem, must be taken
into account in high precision CPT symmetry tests.

The last remark. Within the standard theory of neutral meson
complexes all evolution equations are derived from the
Schr\"{o}dinger Equation (\ref{Schr}) using more or less accurate
approximations (see \cite{LOY1} --- \cite{data}, \cite{bell} ---
\cite{tsai} and so on). This means that there is a consensus that
the Schr\"{o}dinger Equation describes correctly time evolution in
such systems. So, if we adopt this opinion and assume that the
Schr\"{o}dinger Equation describes correctly real properties of the
varying in time processes in the the systems, e.g., containing
neutral meson complex as a subsystem, then we  must also accept all
rigorous consequences of such an assumption. The main conclusions of
this paper are consequences of this type.

\end{document}